# Refining Au/Sb alloyed ohmic contacts in undoped Si/SiGe strained quantum wells

LuckyDonald L Kynshi[1], Umang Soni[1], Chithra H Sharma[2,3], Shengqiang Zhou[4] and, Madhu Thalakulam[1,*]

[1] *Department of Physics, Indian Institute of Science Education and Research, Thiruvananthapuram, Kerala 695551, India*

[2] *Institut für Experimentelle und Angewandte Physik, Christian-Albrechts-Universität zu Kiel, 24098 Kiel, Germany*

[3] *Center for Hybrid Nanostructures, Universität Hamburg, Luruper Chaussee 149, 22761 Hamburg, 22761 Germany*

[4] *Helmholtz-Zentrum Dresden-Rossendorf, Institute of Ion Beam Physics and Materials Research, Bautzner Landstraße 400, 01328 Dresden, Germany*

## ABSTRACT

Shallow undoped Si/SiGe quantum wells are the leading platforms for hosting quantum processors based on spin-qubits. The ohmic contacts to the electron gas in these systems are accomplished by ion-implantation technique since the conventional Au/Sb alloyed contacts present a rough surface consisting of sharp islands and pits. These sharp protrusions cause electrical discharge across the gate-dielectric between the ohmic contacts and the accumulation-gates causing device break-down. A clear understanding of the surface morphology, elemental, compositional and electrical characterization of the alloyed region would enable one to engineer a smoother post alloyed surface. In this work, we find that the rough surface morphology is a cumulative effect of the Au/Si eutectic reaction and the threading dislocations inherent in the heterostructure. The structural, elemental, and chemical-state analysis show that the inverted pyramidal pits are resulting from the enhanced Au/Si eutectic reaction at the threading dislocations stemming from the heterostructure interface, while, the sharp protrusions causing accumulation gate-leakage are gold-rich precipitations. The protrusions are removed using an aqua regia treatment prior to the deposition of the gate-oxide and gate electrode. Exploiting a Hall bar device, we analyse the mobility and carrier concentration of the undoped Si/SiGe consisting of Au/Sb alloyed contacts down to 1.5 K. The measured mobility $\sim 10^5 cm^2/Vs$ and carrier concentration of $\sim 10^{11}/cm^2$ are comparable to the reported values on similar high-mobility heterostructures confirming the efficacy of our modified Au/Sb alloy technique in accomplishing high-efficiency contacts to undoped Si/SiGe heterostructures.

[*] madhu@iisertvm.ac.in

## I. INTRODUCTION

Gated quantum dots on a two-dimensional electron gas (DEG) such as that on GaAs/AlGaAs or Si/SiGe semiconductor heterostructures have been identified as one of the leading platforms for the realization of solid-state spin qubits [1–3]. While many of the landmark studies in this direction have been carried out on GaAs/AlGaAs [4–14] based systems, the shorter coherence-times, owing to the nuclei hyperfine interaction prompted researchers to explore other platforms such as Si/SiGe. The coherence times on strained silicon quantum wells in a Si/SiGe heterostructure have shown to exceed by many orders compared to that on GaAs/AlGaAs quantum wells [15]. Isotopically purified $^{28}$Si quantum wells, free of any hyperfine interaction [16,17], boast a much longer coherence-time of $\sim 3s$ [18]. The initial experiments on gated quantum dots on Si/SiGe exploited a doped heterostructure [16,19,20]. Few-electron single and double quantum dots [21], tunable tunnel coupling [22], spin blockade [23], etc., have been demonstrated on doped Si/SiGe systems. The ionized dopants in the dopant layer have been identified as a source of potential fluctuations on these systems causing switching noise, device instability, lack of uniformity, and leakage currents in the transport measurements [24]. Recent advancements in the development of undoped enhancement-mode Si/SiGe systems address these issues by eliminating the dopant layer, achieving higher carrier mobilities, lower leakage current, and improved stability [25].

Establishing high carrier injection efficiency electrical contacts to the 2DEG is an essential exercise in these devices. It is usually achieved by degeneratively doping the source and drain regions to minimize the Schottky barrier present at the metal-semiconductor interface [26]. Au/Sb alloyed ohmic contacts are widely used in modulation-doped Si/SiGe quantum wells [27–29]. It involves depositing either a stack of Au/Sb/Au or Au-Sb alloy followed by a rapid temperature annealing at $\sim 400 - 450$ degrees for a few minutes. The mechanism invokes the eutectic reaction between Au and Si, which involves melting and inter mixing of Si and Au at a much lower temperature of $\sim$ 359 °C, resulting in a uniform distribution of Sb in the alloyed region [30]. The rapid thermal annealing cycle results in a very rough post-alloyed surface consisting of micro-meter scale sharp and protrusive islands and pits. A comprehensive study of the surface morphology, elemental compositional analysis of these alloyed region is wanting at large. The sharp protrusions are identified as a major source of electric discharge between the gate electrode and the ohmic contacts across the gate-oxide causing heavy gate leakage in undoped mode devices, making it unsuitable for enhancement mode operation. Owing to this, on undoped Si/SiGe structures, the ion implantation technique is adopted for the formation of highly doped electrical contacts with the 2DEG. Ion-implantation requires a high annealing temperature, ~ 700 $^0$C, which might lead to the interdiffusion of Si and Ge and subsequent Si strain relaxation and development of dislocations in the heterostructure affecting its electrical performance [31]. Additionally, the ion-implanters are limited to advanced device fabrication facilities while for prototyping devices it is desirable to have an affordable and compact choice with quick turnaround times. Though there are a few early studies of Au/Sb alloyed ohmic contacts on enhancement-mode Si/SiGe 2DEGs [32,33], specifically on deeper 2DEGs, rarely it has been used on shallow 2DEGs which are the main platform for hosting spin qubits.

In this work, we (i) develop a recipe, alternate to ion implantation, for the realization of reliable Au/Sb ohmic contacts to shallow undoped Si/SiGe quantum wells operating down to cryogenic temperatures, and (ii) conduct a detailed structural, material and electrical characterizations of the alloyed regions to bridge the gap of our fundamental understanding on the formation of these alloyed regions. The surface morphology is studied using scanning electron microscopy (SEM) and atomic force microscopy (AFM). Elemental analysis is conducted using energy dispersive x-ray spectroscopy (EDS). In addition, we use x-ray photo-electron spectroscopy (XPS) to determine the chemical state of the alloyed regions. We perform an anisotropic wet chemical etching using $CrO_3$+HF to understand the role of the threading dislocations present in the structure in deciding the surface morphology of the post-alloyed regions. We find that the major source of the sharp protruding islands on the annealed regions are gold precipitations resulting from the Au/Sb eutectic reaction. We use an aqua regia assisted wet chemical etching to etch out these structures to smooth and reduce the roughness of the alloyed regions to make it suitable for enhancement mode devices. A device with a Hall geometry is used to investigate the carrier injection efficiency of the alloyed Au-Sb ohmic contacts, carrier concentration, and mobility of the 2DEG under various top-gate voltages. We find that our alloyed contacts perform at par with the ion-implanted contacts down to a temperature of 1.5 K.

## II. SAMPLE FABRICATION

The undoped Si/SiGe heterostructure used in this work are commercially procured (Lawrence Semiconductor Research Laboratory) and hosts an 8 nm wide Silicon quantum well, sandwiched between two $Si_{0.70}Ge_{0.30}$ layers, ~ 50 nm below the wafer surface [Fig. 1 (a)]. The studies discussed in this work are performed on small pieces, ~ $5mm \times 5mm$, cleaved from the wafer. All material characterization studies are done on samples not subjected to any lithographic processes while the transport studies are done on samples subjected to photolithographically defined Ohmic contact regions. The samples are first subjected to a standard RCA cleaning procedure to remove any organic residues before any processing. We employ a thermally deposited Au/Sb/Au stack for the realization of the ohmic contacts. Prior to the deposition of the Au/Sb/Au stack, the samples are treated in buffer oxide etch solution (BOE) for 30 seconds to remove the native oxide. Post-deposition, the samples are annealed at 450 degrees in a forming gas atmosphere [$H_2$(15%) + Ar (85%)] using a homemade rapid thermal annealing (RTA) setup. Post-annealing, the alloyed regions exhibit a rough surface topography consisting of protrusive sharp regions and pits as illustrated in the inset to Fig. 1 (a). SEM image of the post-alloyed surface is shown in Fig. 1 (b) and a magnified view of the surface microstructures is shown in Fig. 1 (c). Though we have investigated the ohmic contact formation of Au/Sb/Au stacks with various ratios, we find that the ohmic contacts down to cryogenics temperatures are established only for 99% Au and 1%Sb composition [33]. For higher Sb fractions, the behaviour of the contacts are very poor and turns non-responsive at lower temperatures. Here, we focus on samples made with three different recipes; Au/Sb/Au with thicknesses 10/4/110nm (labelled S1), 15/6/170nm (labelled S2) and 15/7/240nm (labelled S3). All three recipes are chosen such that a 99:1 mass fraction is maintained between Au and Sb.

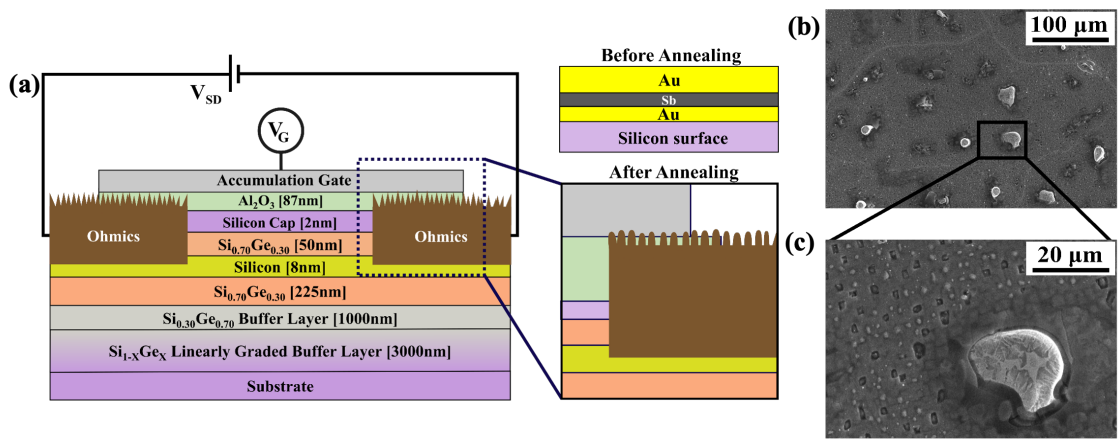

FIG. 1 (a) Schematic diagram of the Si/SiGe heterostructure with Au/Sb alloyed ohmic contacts, $Al_2O_3$ dielectric layer, and top-accumulation gate. $V_{SD}$ is the Source drain bias applied across the ohmic contact and $V_G$ is the accumulation gate voltage. Top-right inset: The order of metallization - Au/Sb/Au - on the silicon surface. A magnified diagram illustrating the island formation after the annealing process. (b) SEM image of Au/Sb contact after an annealing. (c) A magnified SEM image of the annealed region showing the island protruding out of the surface.

Two-probe resistance of the annealed regions are acquired from room temperature down to 4 K. Samples prepared by all three recipes exhibit linear current versus voltage (I-V) characteristics in the entire temperature range. A table summarising the essential characteristics of samples prepared with the three different recipes is provided in the Supplementary Material SI-1. The S2-samples exhibit the lowest resistance among the three and all the analyses discussed in this manuscript are conducted on samples with contacts based on this recipe. Observations pertaining to samples prepared with recipes S1 and S3 are provided in the supplementary information, wherever applicable. The transport studies are conducted on devices with Hall-bar geometry, with a dimension of $105 \mu m \times 405 \mu m$ realized by standard photolithography. Post deposition of the Au/Sb/Au stack and the RTA process, the samples are treated with an aqua regia solution ($HNO_3$: HCl - 1:3) for ten minutes. This results in the removal of the protruding islands in the annealed region and a smoother surface profile minimising the risk of device damage by electrical discharge across the oxide layer. Further, an $Al_2O_3$ layer of $\sim 90 nm$ was deposited using an atomic layer deposition (ALD) system and a 100 nm thick accumulation gate was realized by photolithography followed by Aluminium metallization.

## III. RESULTS & DISCUSSIONS

### A. Surface morphology and elemental composition

The surface morphology of the annealed regions is studied using both AFM and SEM techniques. Figure 2(a) shows the SEM image of an S2-sample (left) while a magnified view of the region marked in black square box is shown on the right. We identify four distinct microstructures on the alloyed region; (i) island, indicated by the orange box in the left panel, (ii) craters, marked by the violet box in the right panel, (iii) bump, outlined by the red box in the right panel, and (iv) a plain smooth region represented by the white spot in Fig. 2 (a) right panel. The resulting anisotropic surface morphology is influenced by the final annealing temperature, cooling rate, stacking faults coming from growth, crystal orientation, and the readjustment in the chemical composition to maintain equilibrium [30,34–36]. The island-like and the bump-like structures result from the anisotropic reaction, the intermixing of gold and silicon, governed by the Au-Si eutectic binary phase relationship [37], and a subsequent precipitation of Au and Si during cooling down process [30]. The crater-like structures are formed in regions where the reaction areas are large and the eutectic alloy do not refill the structure completely, leaving a visible crater on the silicon surface [35]. The samples prepared by the other two recipes, S1 and S3, exhibit a similar surface morphology as shown in the Supplementary Material SI-2.

We extract a three-dimensional profile of the surface morphology using AFM to characterize the bump and the crater structures, as shown in Fig. 2 (b). The surface profile confirms the development of craters and bumps post-alloying. The protrusive island region represented by the orange box in Fig. 2 (a) left panel is too high ($\sim 2-5\mu m$) to acquire the height profile using the AFM. The line-profiles in Fig.2 (c) shows that the height of the small-bumps(red-box) and the depth of the craters (violet-box) are a few hundred nanometres ($\sim 380 nm$) in dimension. The extracted depth of the craters suggests that they originate from the $Si_{0.7}Ge_{0.3}$ barrier layer grown on top of the buffer layer [Fig. 1(a)]. Presence of similar inverted pyramidal pits have been reported on strained silicon quantum wells, and the origin has been attributed to screw and misfit dislocations nucleating at the Si/SiGe interface [38–40].

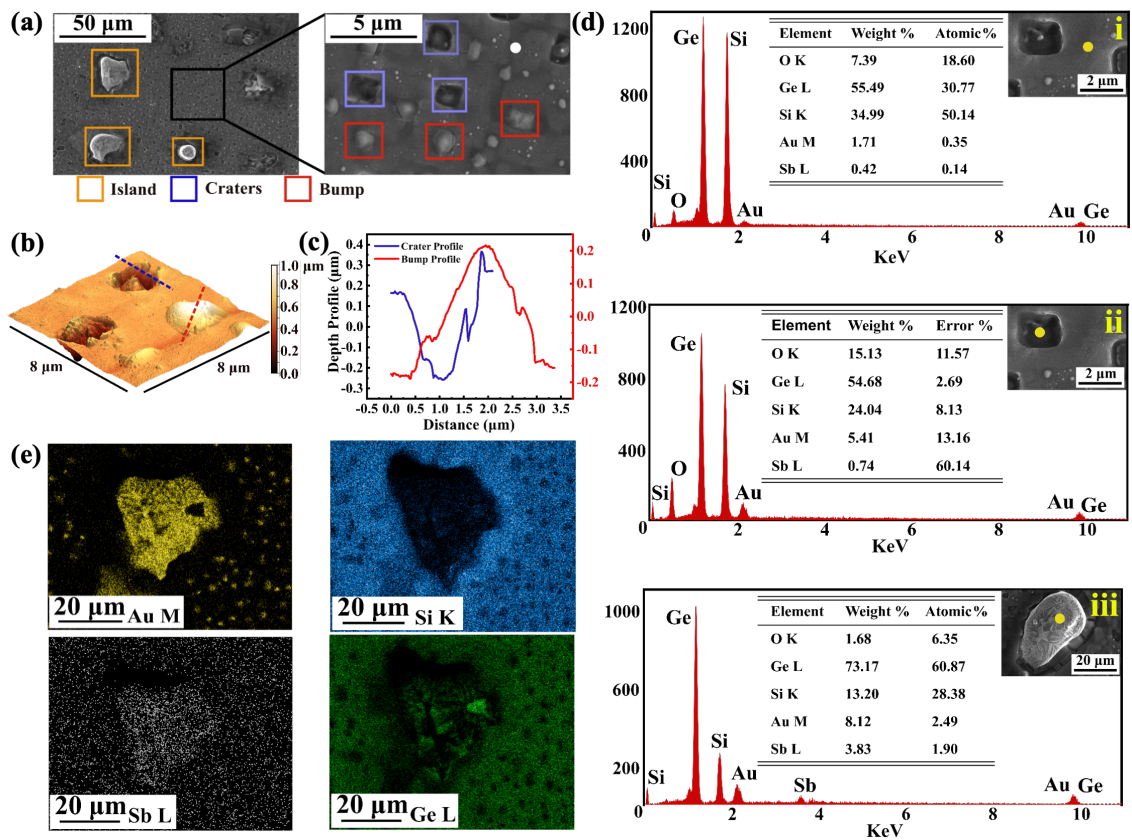

FIG.2. (a) SEM image of the annealed region at 50μm(left) and 5μm(right) resolutions. Various post-alloying microstructures island (orange-square), rectangular craters (blue squares), small bumps (red square), and relatively plain regions (white dot). (b) The 3D AFM surface topography shows the crater and the bump-like structure. (c) The line cut along the crater (blue), and bump-like(red) structures. (d) EDS spectra taken at a specific position marked as **i** (plain), **ii** (crater), and **iii** (Island). (e) Elemental mapping showing the distribution of Au (Au M), Si (Si K), Sb (Sb L), and Ge (Ge L) taken on a region represented by orange square of (a) left panel.

We acquire point EDS by focussing the electron beam to extract the elemental composition of different structures in the annealed region. EDS spectra for the plain region is shown on the top panel (marked **i**), while that for the square crater is shown in the middle (marked **ii**), and the island is shown in the bottom (marked **iii**). The chosen positions for recording the EDS spectra are represented by the yellow spots on the SEM images shown in the respective insets in Fig. 2(d). The spectrum of all three regions shows the presence of Au, Si, Sb, and Ge. The variation of the elemental atomic percentage shows that Au is highest in the island regions (2.49%), then in a square crater (1.06 %), and finally, the lowest in a plain region (0.35%). While, for Si, which is highest on the plain (50.14%) and lowest on the island regions (28.38%). The atomic percentage suggests that a significant amount of gold element is agglomerated in the craters and islands during the annealing process, confirming that the reaction of gold and silicon dominantly occurs in these regions, while it is less prevalent in plain regions. Antimony is detected in all three regions, suggesting that antimony is distributed everywhere, although there is a higher concentration in the crater and the island region. Additionally, the surface exhibits an oxygen content, indicating that some fraction of the silicon atoms react with the ambient air. This also suggests that the silicon atoms are diffusing out of the gold layer, confirming the intermixing of Au and Si during the annealing process. The formation of oxide layers during annealing are also reported in the past and attributed to the diffusion of silicon through the Au layer [41]. S3 samples also exhibit similar behaviour; the details are provided in the supplementary material SI-2.

We conduct elemental mapping to study the distribution of different elements over a larger area of on the annealed region, as shown in Fig. 2 (e). The Au mapping (Au M, top left panel) reveals that the island and the crater regions exhibit a higher distribution of gold while it is less in the plain areas. Sb mapping (Sb L, bottom left panel) exhibits a uniform distribution across the scan window, while a marginally higher concentration is observed at the crater and island regions. The top-right and bottom-right panels in Fig. 2 (e) shows the elemental mapping of Si (Si K) and Ge (Ge L) respectively suggesting these elements are uniformly distributed, as expected, except on the island region. This confirms that even for a larger area, our analysis suggests efficient Au-Si reactions happen more favourably in regions containing the precipitates, whereas antimony, a dopant to silicon, is distributed in the whole region.

The EDS studies suggest that the protrusive island-like structures on the alloyed regions are gold-rich. We treat the sample with aqua regia solution (Hcl: $HNO_3$ - 3:1) to etch these protrusive structures to reduce the island's height and make the surface smoother. Fig. 3(a) and (b) show SEM images of the surface before and after the aqua regia treatment, respectively. Figure 3(c) displays the AFM image on one section of the etched islands after an aqua regia treatment. We see that aqua regia removes the island-like structures present on the surface, leaving behind an irregularly shaped pit. The absence of larger island-like structures could significantly reduce the possibility of a gate leakage on our devices.

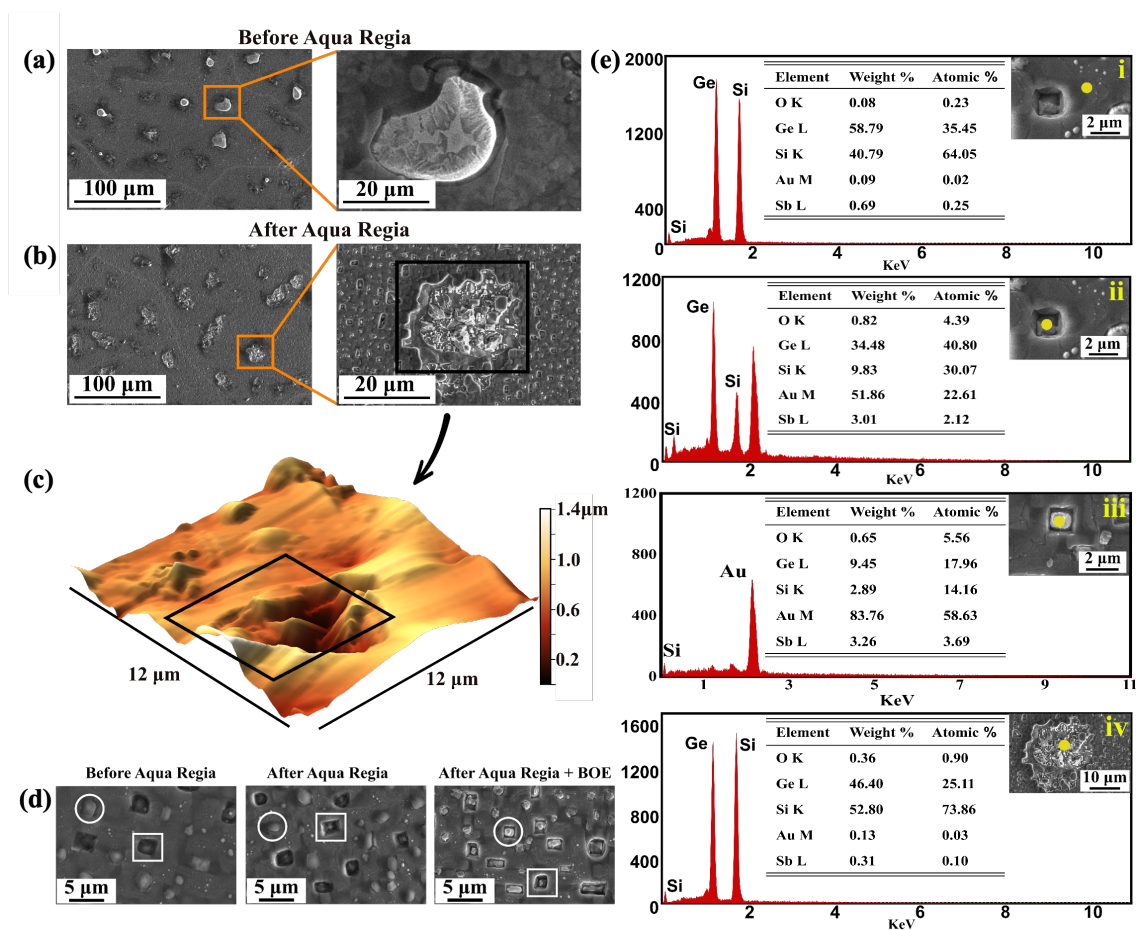

FIG. 3. SEM images of the Au/Sb alloyed region before (a) and after (b) treating with the Aqa Regia at 100μm(left panel) and 20μm(right panel) resolutions. (c) AFM image of the pit created after the gold-rich island has been etched by Aqua Regia. (d) SEM images of the bump like structures (represented by white circle) which is not effectively removed after aqua regia (middle panel) nevertheless removed following the BOE treatment (right panel) revealing a rectangular crater. (e) Point EDS spectrum of the alloyed regions acquired after aqua regia followed by BOE treatment. The SEM images of the respective regions are shown in the insets; plain region (marked i), crater (marked ii), the crater uncovered after the removal of bump-like structures (marked iii), and pits (marked iv) left behind by gold-rich island structures. The position of the EDS data taken is represented by a yellow spot on the respective SEM images.

Fig. 3 (d) shows SEM images of the sample before aqua regia treatment (left) and post-aqua regia treatment (middle). We find that the small bumps, marked in circles, are not much affected by the aqua regia treatment, as shown in Fig. 3(d) middle panel. However, these structures are removed after a treatment with a 6:1 buffered oxide etchant (BOE), leaving a square crater, as shown in Fig. 3 (d) right panel. However, the morphology of the originally present craters, marked in a square box, is not affected by any of these treatments. There is a clear difference between the sharp-island and the small-bump structures. A bump-like structure leaves an inverted pyramidal crater behind after both the aqua regia and BOE treatment [white circles, Fig. 3d (right panel)] whereas an island leaves an irregularly shaped pit structure after the aqua regia treatment [Fig. 2(c)]. S3-sample also reveals a similar surface morphology after the aqua regia and BOE treatment shown in the supplementary material SI-3.

Fig. 3(e) displays the point-EDS analysis of the sample following an aqua regia etching followed by the BOE treatment to remove any oxide layer formed during the annealing. We inspect four distinct regions: the plain region marked by (**i**), a crater (**ii**), the crater left by the small bump-like structure after both the aqua regia and the BOE treatments (**iii**), and the pits left-behind by the large islands after aqua regia treatment (**iv**). We find that the presence of gold is minimal in the plain region and also at the locations of the larger islands post the aqua regia and the BOE treatments. However, inside the craters formed after the BOE treatment, a higher concentration of Au is observed, (**iii**) in Fig. 3(e). The higher concentration of Au may be attributed to the small bump at the centre of the crater containing an Au-Si core covered with a Si-rich layer, which remains unaffected during gold etching. However, following the BOE treatment, the Si atoms are removed, resulting in a crater revealing the gold atoms. EDS studies conducted on an S3 sample displaying similar behaviour is given in Supplementary Material SI-3. We note here that a higher atomic percentage of Sb is present in the crater and island regions, indicating that these areas are heavily doped compared to the surrounding plain region. In addition, the sample shows a lower oxygen content, which we believe is due to the removal of oxide formation by the BOE before acquiring the EDS spectra.

## B. Role of threading dislocations

Owing to the lattice mismatch, Si/SiGe heterostructures are known to possess threading dislocations originating from the heterostructure interfaces [38,40]. These threading dislocations in the heterostructure can trigger inverted pyramidal crater formation. To verify this, we form similar Au/Sb annealed contacts on an intrinsic silicon substrate. Post-annealing SEM image is displayed in the Supplementary Material SI-4. Though image shows prominent island precipitation, we do not find any signatures of craters on the sample suggesting that the dislocations present in the heterostructure formation may be the main factor contributing to the observed square craters observed.

To shine more light onto the appearance of the inverted pyramidal craters post annealing, we inspect the native dislocation density on the Si/SiGe heterostructure with the help of an anisotropic etching procedure using a 2:1 mixture of HF and chromic acid (1M $CrO_3$) [34]. The solution etches faster the defective area, leaving behind a pit or crater, while the defect-free surface remains less affected [42]. The SEM images of the Si/SiGe heterostructure samples, not subject to any metallization and annealing, before and after the $HF + CrO_3$ treatment are shown in Fig. 4(a) and (b) respectively. Post-etching, we observe the appearance of craters of ~1 − 2 $\mu m$ in lateral dimension. The AFM surface profile and the 3D profile shown in Fig. 4 (c) and (d) clearly show a distribution of square craters resembling those observed in Fig. 2 (a) suggesting that these craters are caused by the interfacial strain, and the resulting threading and misfit dislocations [38,40,43]. The extracted depth of these craters and also that of the craters appeared after annealing is of the same order, ~200nm to 300nm. The enhanced eutectic reaction at the dislocations could play a major role in the formation of craters above the eutectic temperature during the annealing process of Au and Si. For comparison, we inspect the surface morphology post $HF + CrO_3$ of an intrinsic silicon sample. The SEM image shown in Fig. 4 (f) shows no craters or etch pits even after a treatment of 25 min with the etchant.

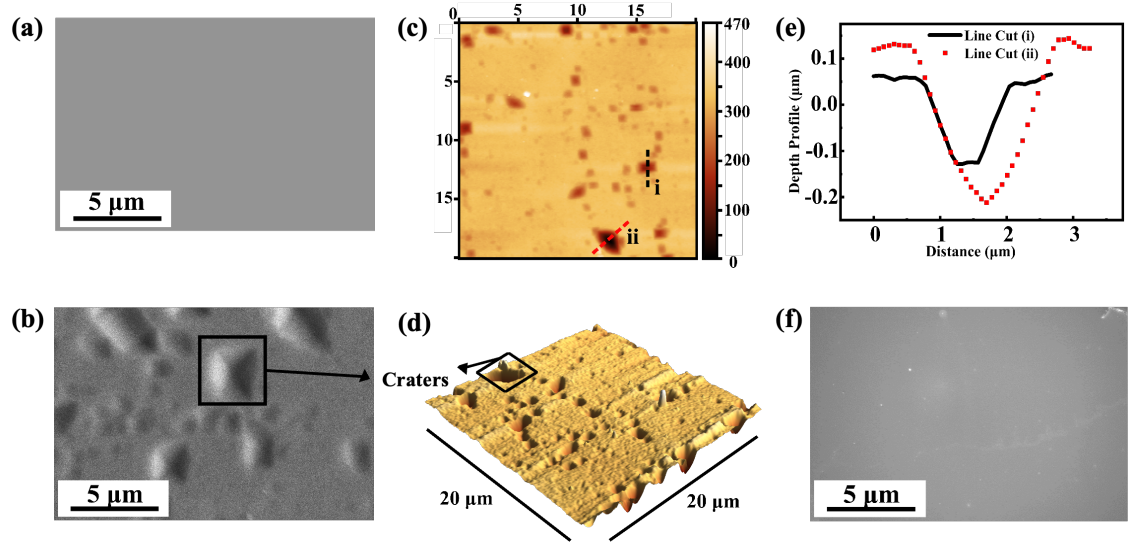

FIG. 4. (a) & (b) The SEM images of the Si/SiGe wafer before (a) and after (b)the $CrO_3$ +HF etching process. (c) & (d) Post $CrO_3$ +HF etching 2D (c)and 3D(d) surface morphology of the samples. (e) Line-profiles extracted for the etch-pits marked i and ii in (c). (f) SEM image of intrinsic silicon after HF + $CrO_3$ treatment showing no signs of craters.

## C. CHEMICAL COMPOSITION

The chemical states of elements in the samples before and after aqua regia treatment are analyzed using XPS. The XPS spectra are carbon corrected to 284.8 eV, and the baseline correction is done using Shirley's method. We find that after annealing, silicon prominently forms oxides in different oxidation states. Figure 5(a) shows the observed $Si^{2+}$ and $Si^{4+}$ species forming Si-O and $SiO_2$ and a sub-oxide state with a $Si^{3+}$ oxidation state [44]. The formation of silicon oxide is also observed from the Sb 3d and O 1s spectra in Fig. 5(b), showing peaks corresponding to $O^{2-}$ states, thus confirming the formation of silicon oxide on the surface. Antimony oxide was also observed in the $Sb^{3+}$ state and $Sb^{5+}$, which can be identified as $Sb_2O_3$ and $Sb_2O_5$ respectively. The Au spectra in Fig. 5(c) show two prominent peaks corresponding to the Au $4f_{7/2-5/2}$ doublets with an energy splitting of 3.7eV. The Au $4f_{7/2}$ is deconvoluted into two peaks at 84.3eV assigned to $Au^0$ and 85.25eV to Gold Silicide $Au^+$ ($Au_xSi_y$). The formation of gold silicides near and above the eutectic temperature, similar to what has been observed in this report has also been reported elsewhere [45–48].

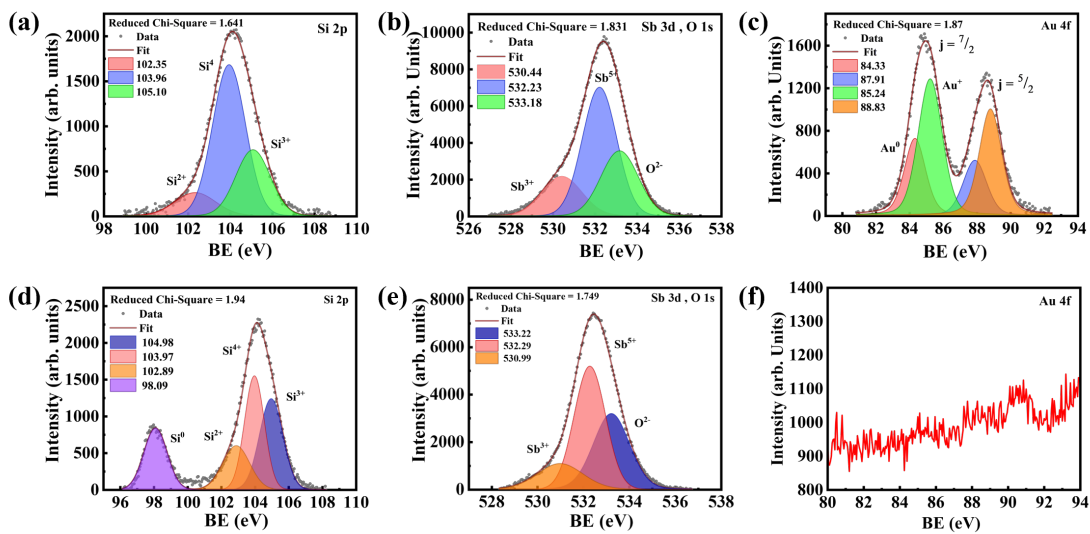

**FIG. 5.** XPS Spectra showing the chemical states of various elements observed before and after aqua regia treatment of annealed samples. (a)-(c) Spectra recorded before aqua regia treatment for Si 2p, Sb 3d, O 1s and Au4f respectively. (d)-(e) post-aqua regia Si 2p, Sb 3d, O 1s and Au4f spectra respectively.

The XPS spectra after the aqua regia solution treatment are shown in Fig. 5(d-f). The Si and the Sb spectra remains similar to that before the aqua regia treatment, Fig. 5 (d) and (e) respectively. Additional peaks are observed at 98 eV, corresponding to the elemental silicon $Si^0$. The absence of $Au^0$ and $Au^+$ after aqua regia treatment confirms the removal of gold silicides present on the surface.

### D. Electrical Characterization

The conductivity of the annealed regions prepared using all three recipes down to a temperature of 4 K was studied. All samples show a linear two-probe I-V characteristics as shown in the Supplementary Material SI-5. S2 samples show the lowest 2-probe resistance

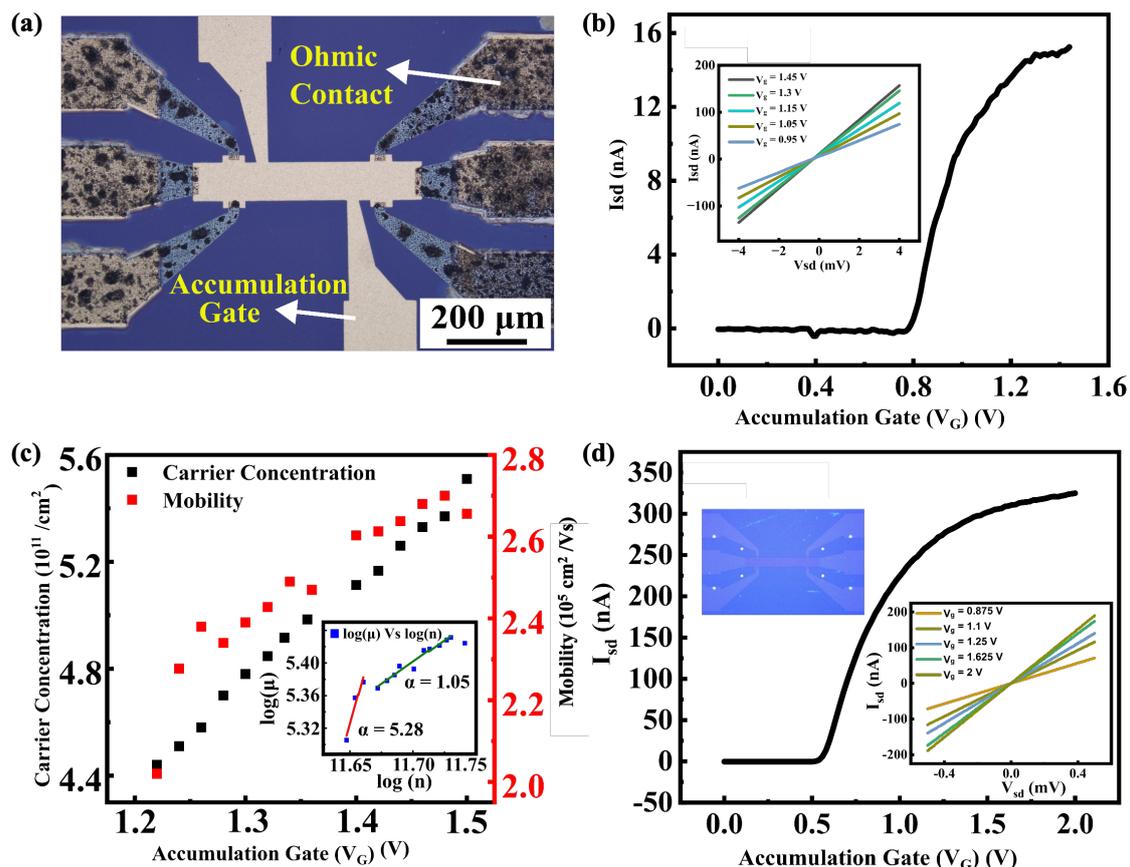

FIG. 6 (a) Optical image of a Hall bar device consisting of Au/Sb alloyed ohmic contacts. (b) Transfer characteristics of the Hall bar exhibiting a turn-on voltage of ~ 0.8V. The inset shows two-probe linear current-voltage characteristics for various top-gate voltages (c) The carrier concentration(left) and mobility(right) versus the top-gate. Inset: log-plot of mobility $\mu$ versus carrier concentration $n$. $\alpha$ is the extracted slope. (d) Transfer characteristic for a Hall-bar (shown in the top-left inset) consisting of ion implanted electrical contacts. Bottom-right inset: two-probe linear current-voltage characteristics for various top-gate voltages.

(~ 2 $k\Omega$). Given the reduced resistance of contacts with S2 recipe, we use the same to make electrical contacts to an undoped Si/SiGe Hall bar to inspect the efficiency of the contacts.

Fig. 6(a) shows a Hall bar device consisting of six ohmic contacts prepared by the recipe S2. The transfer characteristic obtained with a source-drain bias of 0.1 mV of the device obtained at a temperature of 1.5 K is shown in Fig. 6 (b). The two-probe I-V characteristics, acquired for various top-gate voltages, at 1.5 K, shown in the inset to Fig. 6 (b), exhibit an excellent linearity confirming the ohmic nature of the contacts. The device turns on at a top-gate voltage of ~$0.8V$ and the transfer characteristics are similar to what has been reported in the literature [49,50]. A four-probe I-V characteristics of the hall bar at 1.5 K is shown in the Supplementary Material SI-5.

The carrier concentration and mobility of the device for various top-gate voltages are measured using low-field Hall effect measurement at 1.5 K. The black squares in Fig. 6 (c) show a plot of the carrier concentration versus top-gate voltage (left), whereas the extracted mobility for the same range of top-gate voltage is shown in red squares (right). The carrier concentration increases linearly with the top gate voltage, and the calculated carrier concentration is found to be in the $10^{11}/cm^2$ range. The Hall mobility also rises with the top gate voltage, reaching a maximum mobility of ~$2.7 \times 10^5 cm^2/Vs$. Mobility-limiting processes in undoped Si/SiGe heterostructures can be identified using the value of the power-law exponent α of the relation $\mu \propto n^\alpha$ [37]. The inset to Fig. 6(c) shows a log-log plot of mobility against the carrier concentration. At a higher value of the top gate voltage, greater than ~$1.2V$, the electron density increases, promoting electron self-screening from the oxide-interface, and thus enhancing mobility. The value of $\alpha$ extracted is ~1.05, which is attributed to the scattering mechanism due to the remote charges trapped at the $Al_2O_3$/Si interface, limiting the mobility [25,51]. At a lower value of the gate voltages, lower than 1.2V, the slope changes from 1.05 to 5.28 which is a signature of metal-insulation transition in the channel.

The transport characteristics of the Au/Sb ohmic contacted samples are comparable to those consisting of ion implanted contact regions [51]. Fig. 6 (d) shows the current versus top-gate voltage for a Hall of the same dimension with phosphorous ion-implanted contacts. The inset shows a two-probe I-V characteristics of the sample. Both the contact resistance and the turn-on characteristics are similar to that we obtained for samples with Au/Sb alloyed contacts, indicating that our Au/Sb ohmic contact works similarly to those realized by phosphorus ion implantation technique. The obtained results suggest that other than ion implantation, Au/Sb alloy can be implemented as an ohmic contact for the fabrication of mesoscopic devices on shallow undoped Si/SiGe heterostructure.

## IV. Conclusions

This work addresses the challenges of implementing Au/Sb as an ohmic contact to an undoped Si/SiGe system and presents a simple recipe to establish high-efficiency contacts to realize shallow enhancement-mode devices. We find that contacts made with a lower Au/Sb ratio freeze out at cryogenic temperature while, those with higher thickness of Au/Sb, maintaining a mass-fraction of 99:1 shows an ohmic behaviour.  We believe that, increasing the thickness of Au/Sb reduces the alloy resistance, resulting in a more efficient eutectic reaction and a uniform distribution of antimony in the contact regions. A comprehensive study of the alloyed region, consisting of morphological, elemental, and compositional characterizations reveal the details of the microstructures present in the alloyed region. The alloying process results in the intermixing of gold and silicon, which forms gold silicides of various compositions and also silicon oxide on the surface while uniformly distributing Sb, the dopant.  The annealing process leaves the surface rough with protruding gold-rich, island like, structures and also inverted rectangular pyramidal pits. With the help of a defect-etching process, we find that the pyramidal craters are originating due to the activated eutectic reaction at the threading dislocations sites originating from the Si/SiGe heterostructure interface. The large island-like structures are results of the gold-rich precipitations post alloying which are known to cause electric discharge across the dielectric layer between the gate and the source/drain contacts. With the help of an Aqua Regia treatment, we remove the protruding gold-rich structures making these alloyed regions smooth and compatible to realize enhancement mode devices. Electrical transport measurements down to a temperature of 1.5 K conducted on devices with Hall bar geometry reveals a carrier concentration in the range of ~ $10^{11}/cm^2$ and a mobility of $\sim 10^5 cm^2/Vs$ suitable for solid-state quantum devices applications. We also confirm that the electrical characteristics of the alloyed contacts are similar to those realized by phosphorous ion-implantation, suggesting that this method can serve as an alternative method for the realization of quantum devices on shallow undoped Si/SiGe heterostructures.

## Acknowledgments

MT acknowledges DST, Govt of India, for financial support and Anil Shaji for a critical reading of the manuscript. LLK acknowledges PMRF, Govt of India for fellowship and CeNSE, Indian Institute of Science, Bengaluru, for AFM

## References

[1] N. Wang, S.-M. Wang, R.-Z. Zhang, J.-M. Kang, W.-L. Lu, H.-O. Li, G. Cao, B.-C. Wang, and G.-P. Guo, Pursuing high-fidelity control of spin qubits in natural Si/SiGe quantum dot, Appl. Phys. Lett. **125**, 204002 (2024).

[2] E. Kawakami et al., Gate fidelity and coherence of an electron spin in an Si/SiGe quantum dot with micromagnet, Proc. Natl. Acad. Sci. **113**, 11738 (2016).

[3] X. Xue, T. F. Watson, J. Helsen, D. R. Ward, D. E. Savage, M. G. Lagally, S. N. Coppersmith, M. A. Eriksson, S. Wehner, and L. M. K. Vandersypen, Benchmarking Gate Fidelities in a Si / SiGe Two-Qubit Device, Phys. Rev. X **9**, 021011 (2019).

[4] S. Mondal, G. C. Gardner, J. D. Watson, S. Fallahi, A. Yacoby, and M. J. Manfra, Field-effect-induced two-dimensional electron gas utilizing modulation-doped ohmic contacts, Solid State Commun. **197**, 20 (2014).

[5] Y. Ashlea Alava, D. Q. Wang, C. Chen, D. A. Ritchie, O. Klochan, and A. R. Hamilton, High electron mobility and low noise quantum point contacts in an ultra-shallow all-epitaxial metal gate GaAs/AlxGa1–xAs heterostructure, Appl. Phys. Lett. **119**, 063105 (2021).

[6] M. Kemerink and L. W. Molenkamp, Stochastic Coulomb blockade in a double quantum dot, Appl. Phys. Lett. **65**, 1012 (1994).

[7] F. R. Waugh, M. J. Berry, D. J. Mar, R. M. Westervelt, K. L. Campman, and A. C. Gossard, Single-Electron Charging in Double and Triple Quantum Dots with Tunable Coupling, Phys. Rev. Lett. **75**, 705 (1995).

[8] W. Y. Mak, F. Sfigakis, H. E. Beere, I. Farrer, J. P. Griffiths, G. a. C. Jones, D. A. Ritchie, K. Das Gupta, O. Klochan, and A. R. Hamilton, Ultra-shallow quantum dots in an undoped GaAs/AlGaAs two-dimensional electron gas, Appl. Phys. Lett. **102**, (2013).

[9] W. G. Van Der Wiel, S. De Franceschi, J. M. Elzerman, T. Fujisawa, S. Tarucha, and L. P. Kouwenhoven, Electron transport through double quantum dots, Rev. Mod. Phys. **75**, 1 (2002).

[10] J. R. Petta, A. C. Johnson, J. M. Taylor, E. A. Laird, A. Yacoby, M. D. Lukin, C. M. Marcus, M. P. Hanson, and A. C. Gossard, Coherent Manipulation of Coupled Electron Spins in Semiconductor Quantum Dots, Science **309**, 2180 (2005).

[11] F. H. L. Koppens, C. Buizert, K. J. Tielrooij, I. T. Vink, K. C. Nowack, T. Meunier, L. P. Kouwenhoven, and L. M. K. Vandersypen, Driven coherent oscillations of a single electron spin in a quantum dot, Nature **442**, 766 (2006).

[12] H.-O. Li, G. Cao, M. Xiao, J. You, D. Wei, T. Tu, G.-C. Guo, H.-W. Jiang, and G.-P. Guo, Fabrication and characterization of an undoped GaAs/AlGaAs quantum dot device, J. Appl. Phys. **116**, 174504 (2014).

[13] W. Jang, M.-K. Cho, H. Jang, J. Kim, J. Park, G. Kim, B. Kang, H. Jung, V. Umansky, and D. Kim, Single-Shot Readout of a Driven Hybrid Qubit in a GaAs Double Quantum Dot, Nano Lett. **21**, 4999 (2021).

[14] I. A. Merkulov, Al. L. Efros, and M. Rosen, Electron spin relaxation by nuclei in semiconductor quantum dots, Phys. Rev. B **65**, 205309 (2002).

[15] K. Takeda et al., Optimized electrical control of a Si/SiGe spin qubit in the presence of an induced frequency shift, Npj Quantum Inf. **4**, 54 (2018).

[16] C. B. Simmons, M. Thalakulam, N. Shaji, L. J. Klein, H. Qin, R. H. Blick, D. E. Savage, M. G. Lagally, S. N. Coppersmith, and M. A. Eriksson, Single-electron quantum dot in Si/SiGe with integrated charge-sensing, Appl. Phys. Lett. **91**, 213103 (2007).

[17] A. M. Tyryshkin and others, Electron spin coherence exceeding seconds in high-purity silicon, Nat. Mater. **11**, 143 (2011).

[18] C. B. Simmons et al., Tunable Spin Loading and T 1 of a Silicon Spin Qubit Measured by Single-Shot Readout, Phys. Rev. Lett. **106**, 156804 (2011).

[19] M. Thalakulam, C. B. Simmons, B. J. Van Bael, B. M. Rosemeyer, D. E. Savage, M. G. Lagally, M. Friesen, S. N. Coppersmith, and M. A. Eriksson, Single-shot measurement and tunnel-rate spectroscopy of a Si/SiGe few-electron quantum dot, Phys. Rev. B **84**, 045307 (2011).

[20] C. B. Simmons et al., Pauli spin blockade and lifetime-enhanced transport in a Si/SiGe double quantum dot, Phys. Rev. B **82**, 245312 (2010).

[21] M. Thalakulam, C. B. Simmons, B. M. Rosemeyer, D. E. Savage, M. G. Lagally, M. Friesen, S. N. Coppersmith, and M. A. Eriksson, Fast tunnel rates in Si/SiGe one-electron single and double quantum dots, Appl. Phys. Lett. **96**, 183104 (2010).

[22] C. B. Simmons et al., Charge Sensing and Controllable Tunnel Coupling in a Si/SiGe Double Quantum Dot, Nano Lett. **9**, 3234 (2009).

[23] N. Shaji et al., Spin blockade and lifetime-enhanced transport in a few-electron Si/SiGe double quantum dot, Nat. Phys. **4**, 540 (2008).

[24] K. A. Slinker et al., Quantum dots in Si/SiGe 2DEGs with Schottky top-gated leads, New J. Phys. **7**, 246 (2005).

[25] Y.-H. Su, K.-Y. Chou, Y. Chuang, T.-M. Lu, and J.-Y. Li, Electron mobility enhancement in an undoped Si/SiGe heterostructure by remote carrier screening, J. Appl. Phys. **125**, 235705 (2019).

[26] D. Zhang, G. Yuan, S. Zhao, J. Lu, and J. Luo, Low-thermal-budget n-type ohmic contacts for ultrathin Si/Ge superlattice materials, J. Phys. Appl. Phys. **55**, 355110 (2022).

[27] P. J. Wang, B. S. Meyerson, K. Ismail, F. F. Fang, and J. Nocera, High Mobility Two-Dimensional Electron Gas in Modulation-Doped Si/SiGe Heterostructures, MRS Proc. **220**, 403 (1991).

[28] J. Sailer et al., A Schottky top-gated two-dimensional electron system in a nuclear spin free Si/SiGe heterostructure, Phys. Status Solidi RRL – Rapid Res. Lett. **3**, 61 (2009).

[29] C. Payette, K. Wang, P. J. Koppinen, Y. Dovzhenko, J. C. Sturm, and J. R. Petta, Single charge sensing and transport in double quantum dots fabricated from commercially grown Si/SiGe heterostructures, Appl. Phys. Lett. **100**, 043508 (2012).

[30] J. Q. Liu, C. Wang, T. Zhu, W. J. Wu, J. Fan, and L. C. Tu, Low temperature fabrication and doping concentration analysis of Au/Sb ohmic contacts to n-type Si, AIP Adv. **5**, 117112 (2015).

[31] Y. Xu et al., On-chip Integration of Si/SiGe-based Quantum Dots and Switched-capacitor Circuits, Appl. Phys. Lett. **117**, 144002 (2020).

[32] T. M. Lu, D. C. Tsui, C.-H. Lee, and C. W. Liu, Observation of two-dimensional electron gas in a Si quantum well with mobility of $1.6 \times 10^6$ cm$^2$/Vs, Appl. Phys. Lett. **94**, 182102 (2009).

[33] A. A. Shashkin, V. T. Dolgopolov, S.-H. Huang, and S. V. Kravchenko, Ultra-high mobility two-dimensional electron gas in a SiGe/Si/SiGe quantum well, Appl. Phys. Lett. **106**, (2015).

[34] N. Ferralis, R. Maboudian, and C. Carraro, Temperature-Induced Self-Pinning and Nanolayering of AuSi Eutectic Droplets, J. Am. Chem. Soc. **130**, 2681 (2008).

[35] T. Ye, Z. Song, Y. Du, and Z. Wang, *Reliability of Au-Si Eutectic Bonding*, in *2014 15th International Conference on Electronic Packaging Technology* (2014), pp. 1080–1082.

[36] H. Kato, Eutectic Reactions and Textures of Au–Si Alloy Films on Single-Crystal Silicon, Jpn. J. Appl. Phys. **28**, 953 (1989).

[37] A. Cros and C. Canella, The role of epitaxy in Au-Si eutectic bonding, J. Adhes. Sci. Technol. **5**, 1041 (1991).

[38] A. D. Capewell, Novel Grading of Silicon Germanium for High , Quality Virtual Substrates, (2002).

[39] M. Iwabuchi, K. Mizushima, M. Mizuno, and Y. Kitagawara, Dependence of Epitaxial Layer Defect Morphology on Substrate Particle Contamination of Si Epitaxial Wafer, J. Electrochem. Soc. **147**, 1199 (2000).

[40] D. J. Paul, Si/SiGe heterostructures: from material and physics to devices and circuits, Semicond. Sci. Technol. **19**, R75 (2004).

[41] C.-A. Chang and G. Ottaviani, Outdiffusion of Si through gold films: The effects of Si orientation, gold deposition techniques and rates, and annealing ambients, Appl. Phys. Lett. **44**, 901 (1984).

[42] D. G. Schimmel, Defect Etch for Silicon Evaluation, J. Electrochem. Soc. **126**, 479 (1979).

[43] M. M. Kivambe, D. M. Powell, S. Castellanos, M. A. Jensen, A. E. Morishige, B. Lai, R. Hao, T. S. Ravi, and T. Buonassisi, Characterization of high-quality kerfless epitaxial silicon for solar cells: Defect sources and impact on minority-carrier lifetime, J. Cryst. Growth **483**, 57 (2018).

[44] J. K. Dora, D. Nayak, S. Ghosh, V. Adyam, N. Yedla, and T. K. Kundu, A facile and green synthesis approach to derive highly stable SiOx-hard carbon based nanocomposites for use as the anode in lithium-ion batteries, Sustain. Energy Fuels **4**, 6054 (2020).

[45] N. Sumida and K. Ikeda, Cross-sectional observations of gold-silicon reaction on silicon substrate in situ in the high-voltage electron microscope, Ultramicroscopy **39**, 313 (1991).

[46] P.-H. Chang, G. Berman, and C. C. Shen, Transmission electron microscopy of gold-silicon interactions on the backside of silicon wafers, J. Appl. Phys. **63**, 1473 (1988).

[47] L. L. Liwei Lin, Y.-T. C. Yu-Ting Cheng, and K. N. Khalil Najafi, Formation of Silicon-Gold Eutectic Bond Using Localized Heating Method, Jpn. J. Appl. Phys. **37**, L1412 (1998).

[48] A. K. Green and E. Bauer, Formation, structure, and orientation of gold silicide on gold surfaces, J. Appl. Phys. **47**, 1284 (1976).

[49] T. M. Lu, D. C. Tsui, C.-H. Lee, and C. W. Liu, Observation of two-dimensional electron gas in a Si quantum well with mobility of 1.6×106 cm2/Vs, Appl. Phys. Lett. **94**, 182102 (2009).

[50] D. Zhang, G. Yuan, Y. Liu, Z. Li, L. Song, J. Lu, J. Zhang, J. Zhang, and J. Luo, Gate-controlled hysteresis curves and dual-channel conductivity in an undoped Si/SiGe 2DEG structure, J. Phys. Appl. Phys. **56**, 085302 (2023).

[51]   X. Mi, T. M. Hazard, C. Payette, K. Wang, D. M. Zajac, J. V. Cady, and J. R. Petta, Magnetotransport studies of mobility limiting mechanisms in undoped Si/SiGe heterostructures, Phys. Rev. B **92**, 035304 (2015).

LuckyDonald L Kynshi[1], Umang Soni[1], Chithra H Sharma[2,3], Shengqiang Zhou[4], and Madhu Thalakulam[1,*]

[1] *Department of Physics, Indian Institute of Science Education and Research, Thiruvananthapuram, Kerala 695551, India*
[2] *Institut für Experimentelle und Angewandte Physik, Christian-Albrechts-Universität zu Kiel, 24098 Kiel, Germany*
[3] *Center for Hybrid Nanostructures, Universität Hamburg, Luruper Chaussee 149, 22761 Hamburg, 22761 Germany*
[4] *Helmholtz-Zentrum Dresden-Rossendorf, Institute of Ion Beam Physics and Materials Research, Bautzner Landstraße 400, 01328 Dresden, Germany*

**SI-1**

**Table. I Characteristics of different Au/Sb/Au recipes**

| Recipe | Au/Sb/Au Stack | Annealing Temperature | Resistance | | Behaviour |
|---|---|---|---|---|---|
| | | | Room Temperature | 4 k | |
| S1 | 10/4/110 | 450° C, 4min 30 Sec | 4.7 KΩ | 604 K ohm | Ohmic |
| S2 | 15/6/170 | 450° C, 8 min | 2.76 KΩ | 2.1 K ohm | Ohmic |
| S3 | 15/7/240 | 450° C, 8 min | 3.6 KΩ | 9.93K ohm | Ohmic |

[*] madhu@iisertvm.ac.in

## SI-2. Analysis of the annealed region

Figure SI-2 (a-b) shows the surface morphology of the samples prepared with recipes S1 and S3 after an annealing process. Both samples have a rough morphology consisting of islands, bump like structures, and craters. As discussed in the main text of section III A, the eutectic reaction between gold and silicon at the eutectic temperature of all three recipes is validated. Figure SI-2 (c) displays the point analysis conducted for sample with recipe S3. The elemental weight percentage and atomic weight variation are similar to S2-sample. An island region contains the most gold, followed by bump-like structures, and a plain region contains the least amount. The distribution of antimony is similar to that observed in S2, with a slightly higher weight percentage in the island region. The consistent observation in the surface morphology of all samples and the EDS analysis confirms the repeatability of the reaction between Gold and Silicon in all cases.

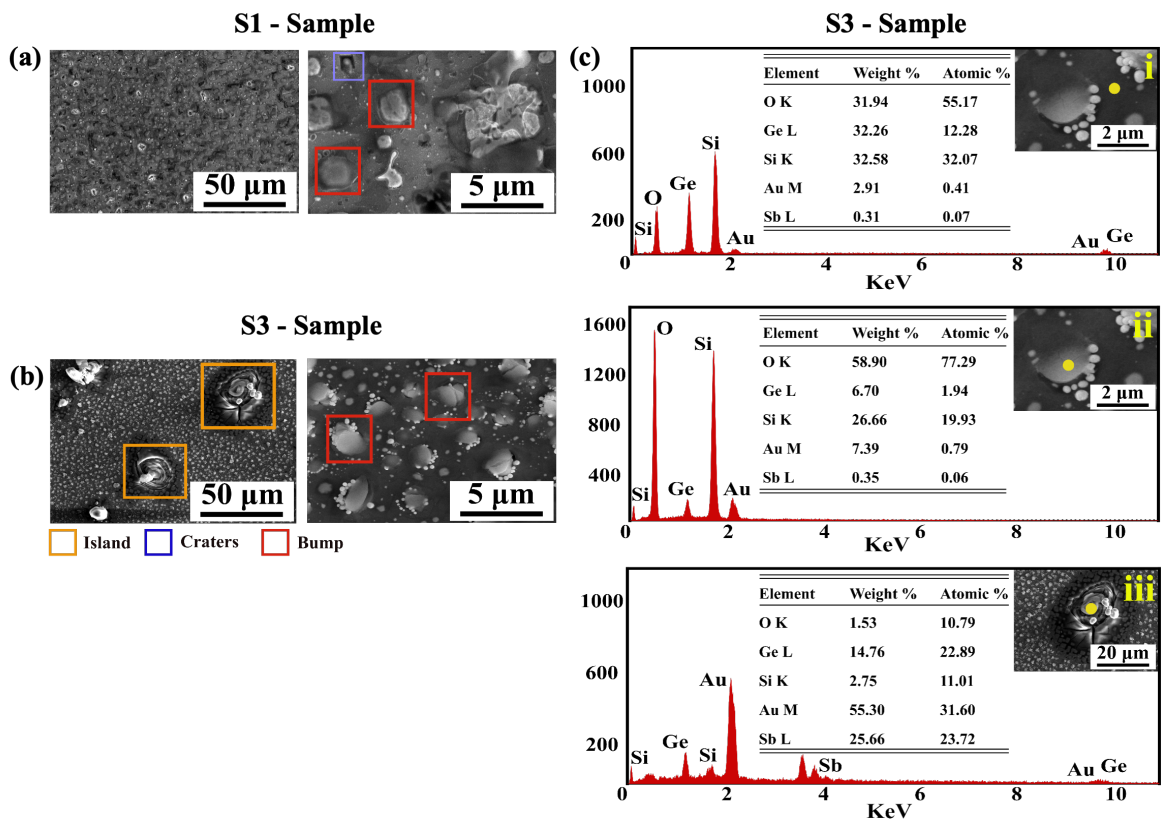

Fig. SI-2 (a) The SEM image of the samples S1 and S3 after an annealing. The morphology contains a bump like structures, island and craters. (b) shows the EDS spectra of a sample S3 taken at a specific position marked as 'i'(plain), 'ii' (crater), 'iii' (Island precipitation). The EDS spectra is taken in the position represented by the yellow spot in the SEM image.

## SI-3. Analysis of the annealed region after aqua regia and BOE treatment

Figure SI-3 (a) shows the SEM images of a sample S3 at various stages before and after etching process. It is clear from the SEM image that the aqua regia removed the 'island' like structures, but did not remove the small bump-like structures on the surface. However, the small bump is removed after the BOE treatment, leaving a crater behind. The EDS spectra in Fig. SI-3 (b) are taken for four different regions represented as 'i' (plain), 'ii' (crater), 'iii' (crater formed after etching the bump like structures), and 'iv' (a pit leaving by the island region). The observed atomic percentage of the elements are similar to S2-samples discussed in the main text.

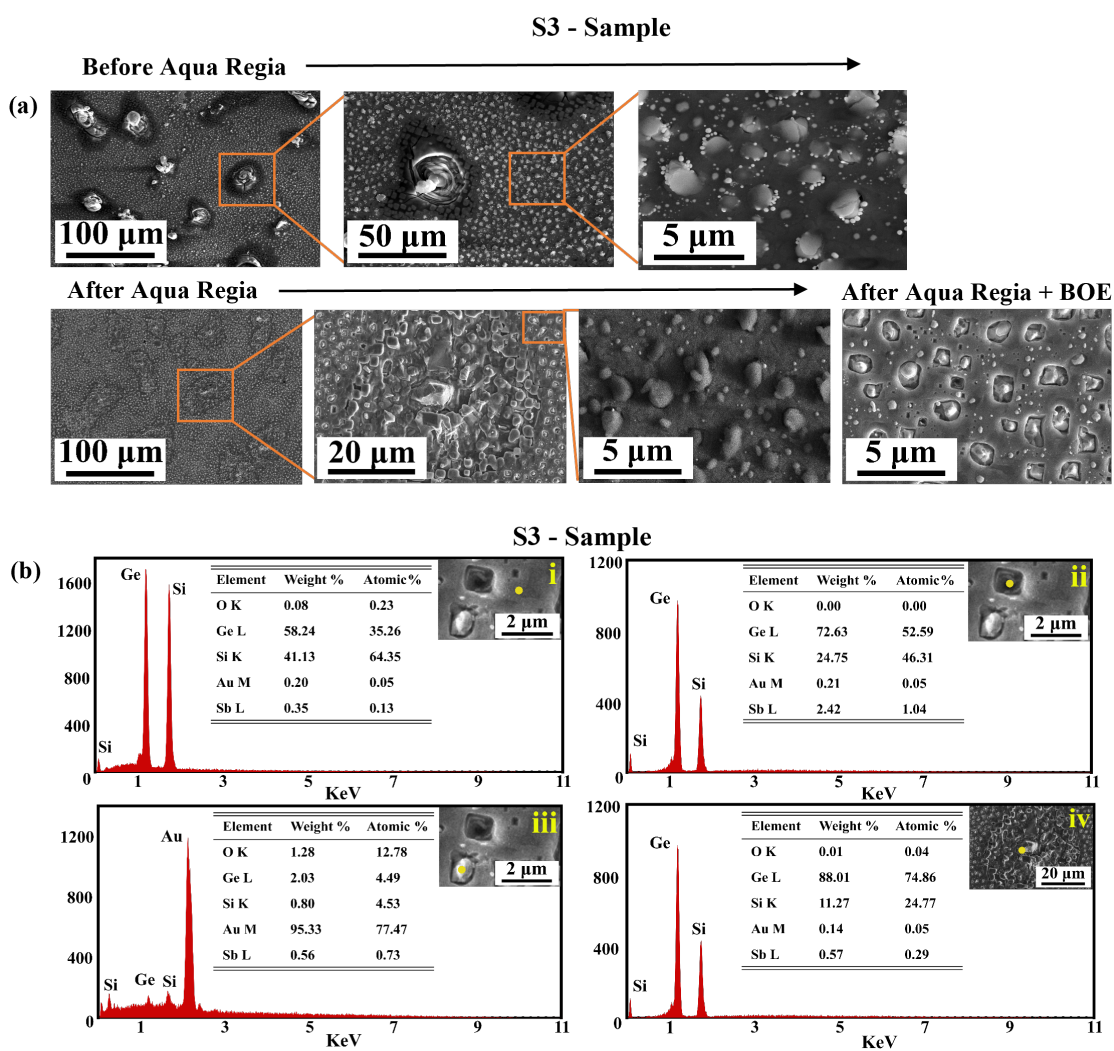

Fig. SI-3 (a) The SEM image of a sample S3 before etching, followed by aqua regia and HF treatment. (b) The EDS analysis of sample S3 after the aqua regia and BOE treatment. The yellow spot represents the position of the taken spectrum represented as 'i'(Plain), 'ii' (crater), 'iii' (crater formed after etching the bump like structures) 'iv' (a pit leaving by the island region)

## SI-4. Analysis of an Intrinsic silicon after an annealing

Figure SI-4 (a) shows an intrinsic silicon SEM image before and after the aqua regia treatment. It is clear that though an island-like structures are observed, however, no distinct crater is observed compared to Si/SiGe discussed in the main text.

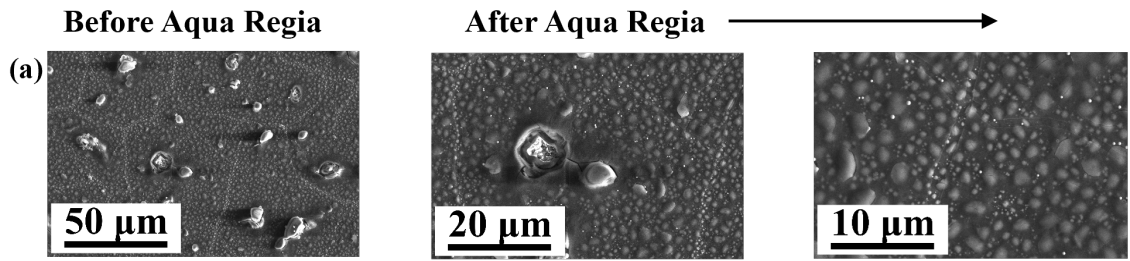

Fig. SI-4 (a) The SEM image of an intrinsic silicon before and after aquaregia shows no signature of craters post annealing.

## SI-5. Electrical Characterization

Figure SI-5 (a-c) illustrates the annealed region's current-voltage (IV) characteristics with recipes S1, S2, and S3 at 4K. All recipes show linear IV characteristics before and after aqua regia treatment. Figure SI-5 (d) is the 4P linear current-voltage characteristics at different top gate voltages of a Hall bar device prepared with S2 recipe, confirming the ohmic behaviour of the ohmic contact at cryogenic temperature.

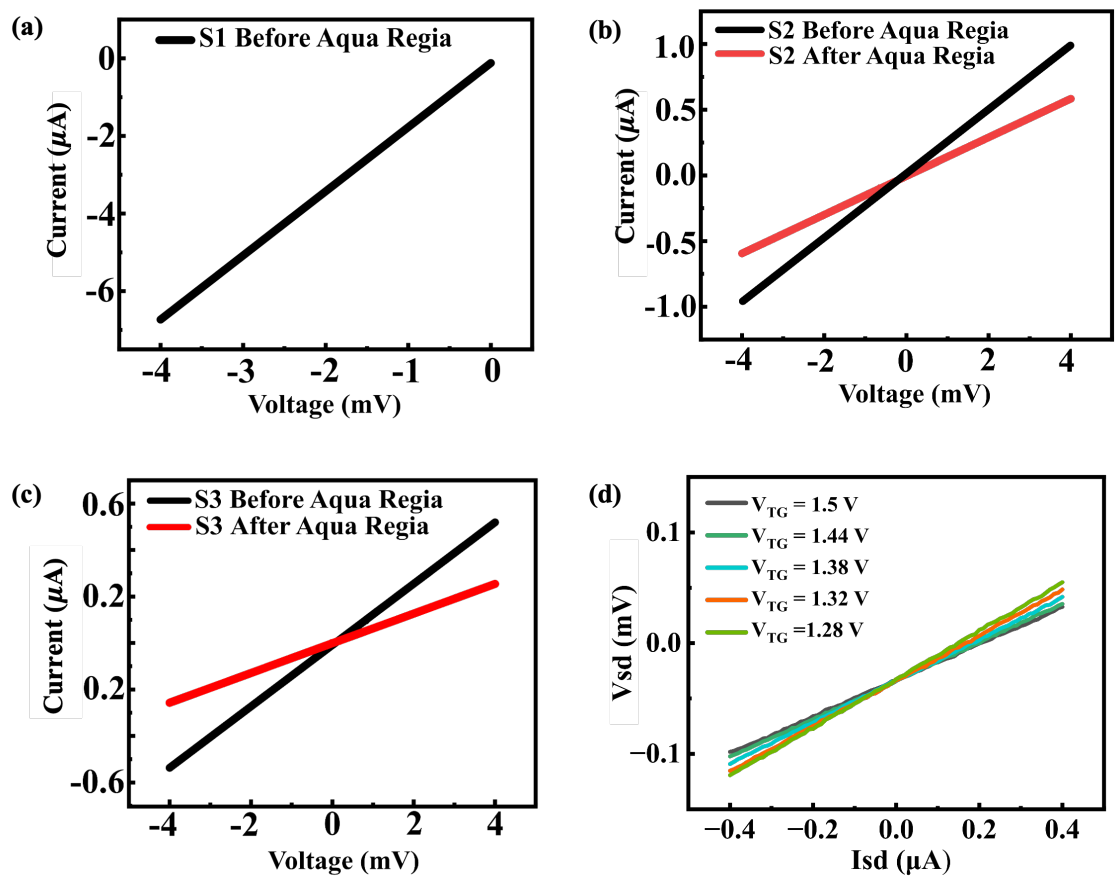

Fig. SI-5 (a) IV characteristics of the annealed region of S1. (b) Linear IV characteristics of S2 before and after aqua regia. (c) Linear IV characteristics of S3 before and after aqua regia (d) Hall bar 4-Probe linear IV characteristic with different top gates at 1.5K prepared with S2 recipe.

# Supplementary material

# Refining Au/Sb alloyed ohmic contacts in undoped Si/SiGe strained quantum wells

LuckyDonald L Kynshi[1], Umang Soni[1], Chithra H Sharma[2,3], Shengqiang Zhou[4], and Madhu Thalakulam[1,*]

[1] *Department of Physics, Indian Institute of Science Education and Research, Thiruvananthapuram, Kerala 695551, India*
[2] *Institut für Experimentelle und Angewandte Physik, Christian-Albrechts-Universität zu Kiel, 24098 Kiel, Germany*
[3] *Center for Hybrid Nanostructures, Universität Hamburg, Luruper Chaussee 149, 22761 Hamburg, 22761 Germany*
[4] *Helmholtz-Zentrum Dresden-Rossendorf, Institute of Ion Beam Physics and Materials Research, Bautzner Landstraße 400, 01328 Dresden, Germany*

**SI-1**

**Table. I Characteristics of different Au/Sb/Au recipes**

| Recipe | Au/Sb/Au Stack | Annealing Temperature | Resistance | | Behaviour |
|---|---|---|---|---|---|
| | | | Room Temperature | 4 k | |
| S1 | 10/4/110 | 450° C, 4min 30 Sec | 4.7 K$\Omega$ | 604 K ohm | Ohmic |
| S2 | 15/6/170 | 450° C, 8 min | 2.76 K$\Omega$ | 2.1 K ohm | Ohmic |
| S3 | 15/7/240 | 450° C, 8 min | 3.6 K$\Omega$ | 9.93K ohm | Ohmic |

[*] madhu@iisertvm.ac.in

## SI-2. Analysis of the annealed region

Figure SI-2 (a-b) shows the surface morphology of the samples prepared with recipes S1 and S3 after an annealing process. Both samples have a rough morphology consisting of islands, bump like structures, and craters. As discussed in the main text of section III A, the eutectic reaction between gold and silicon at the eutectic temperature of all three recipes is validated. Figure SI-2 (c) displays the point analysis conducted for sample with recipe S3. The elemental weight percentage and atomic weight variation are similar to S2-sample. An island region contains the most gold, followed by bump-like structures, and a plain region contains the least amount. The distribution of antimony is similar to that observed in S2, with a slightly higher weight percentage in the island region. The consistent observation in the surface morphology of all samples and the EDS analysis confirms the repeatability of the reaction between Gold and Silicon in all cases.

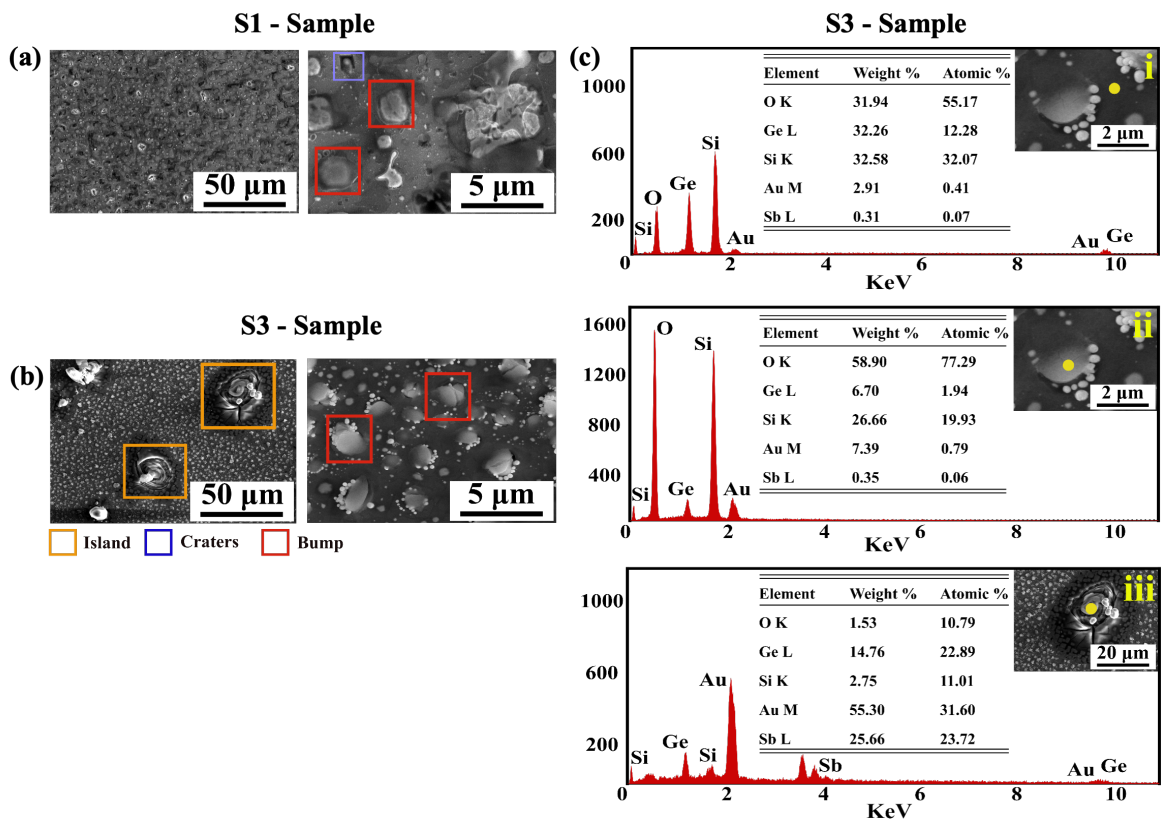

Fig. SI-2 (a) The SEM image of the samples S1 and S3 after an annealing. The morphology contains a bump like structures, island and craters. (b) shows the EDS spectra of a sample S3 taken at a specific position marked as 'i'(plain), 'ii' (crater), 'iii' (Island precipitation). The EDS spectra is taken in the position represented by the yellow spot in the SEM image.

## SI-3. Analysis of the annealed region after aqua regia and BOE treatment

Figure SI-3 (a) shows the SEM images of a sample S3 at various stages before and after etching process. It is clear from the SEM image that the aqua regia removed the 'island' like structures, but did not remove the small bump-like structures on the surface. However, the small bump is removed after the BOE treatment, leaving a crater behind. The EDS spectra in Fig. SI-3 (b) are taken for four different regions represented as 'i' (plain), 'ii' (crater), 'iii' (crater formed after etching the bump like structures), and 'iv' (a pit leaving by the island region). The observed atomic percentage of the elements are similar to S2-samples discussed in the main text.

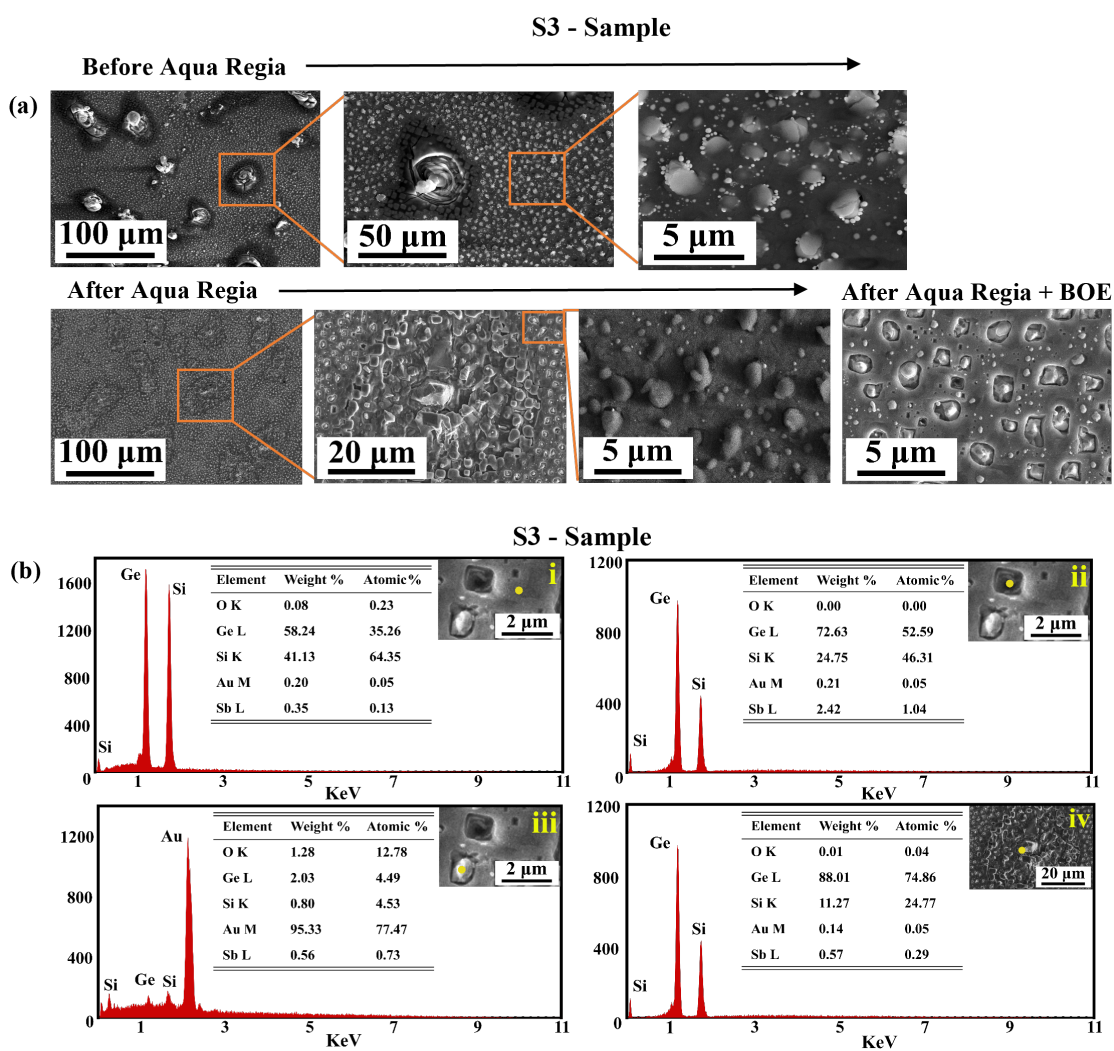

Fig. SI-3 (a) The SEM image of a sample S3 before etching, followed by aqua regia and HF treatment. (b) The EDS analysis of sample S3 after the aqua regia and BOE treatment. The yellow spot represents the position of the taken spectrum represented as 'i'(Plain), 'ii' (crater), 'iii' (crater formed after etching the bump like structures) 'iv' (a pit leaving by the island region)

## SI-4. Analysis of an Intrinsic silicon after an annealing

Figure SI-4 (a) shows an intrinsic silicon SEM image before and after the aqua regia treatment. It is clear that though an island-like structures are observed, however, no distinct crater is observed compared to Si/SiGe discussed in the main text.

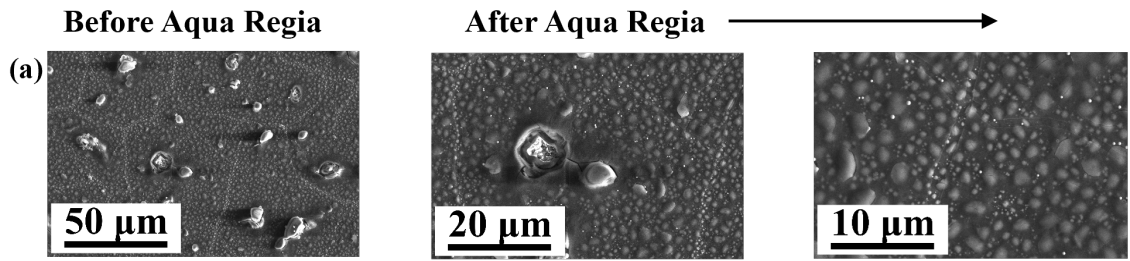

Fig. SI-4 (a) The SEM image of an intrinsic silicon before and after aquaregia shows no signature of craters post annealing.

## SI-5. Electrical Characterization

Figure SI-5 (a-c) illustrates the annealed region's current-voltage (IV) characteristics with recipes S1, S2, and S3 at 4K. All recipes show linear IV characteristics before and after aqua regia treatment. Figure SI-5 (d) is the 4P linear current-voltage characteristics at different top gate voltages of a Hall bar device prepared with S2 recipe, confirming the ohmic behaviour of the ohmic contact at cryogenic temperature.

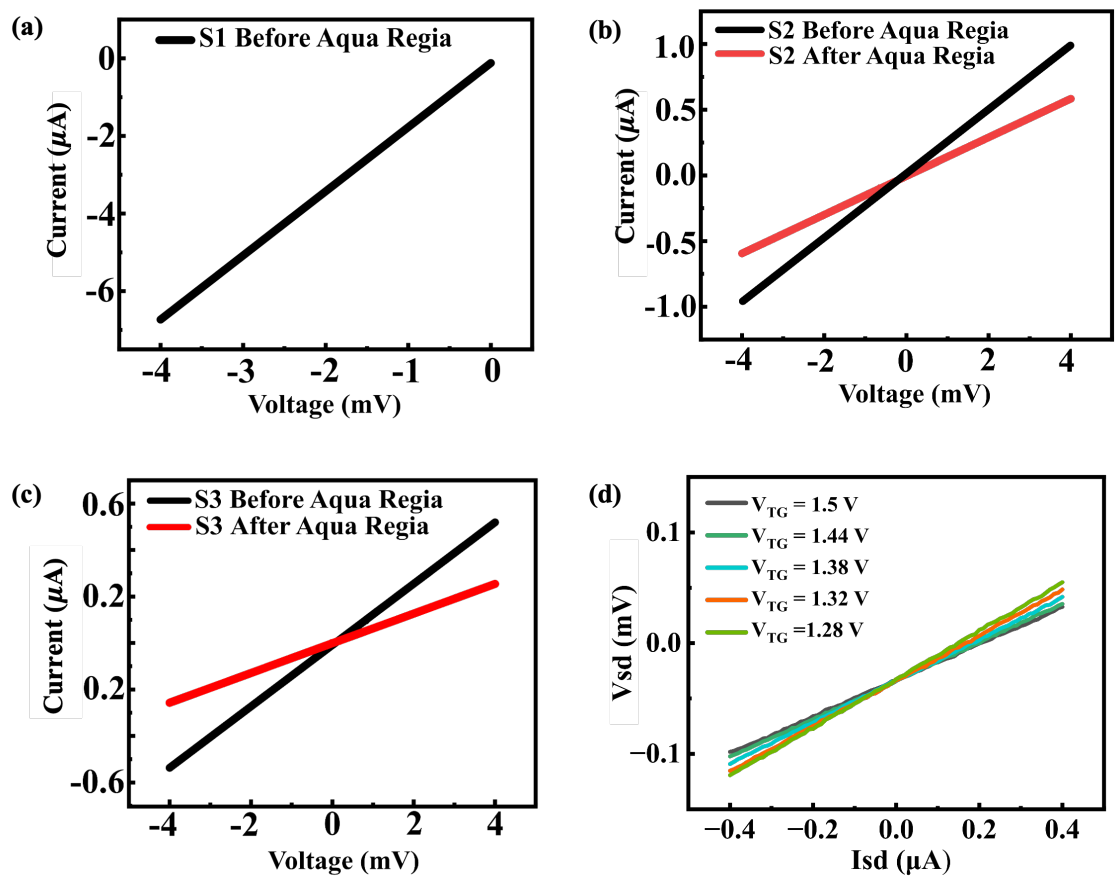

Fig. SI-5 (a) IV characteristics of the annealed region of S1. (b) Linear IV characteristics of S2 before and after aqua regia. (c) Linear IV characteristics of S3 before and after aqua regia (d) Hall bar 4-Probe linear IV characteristic with different top gates at 1.5K prepared with S2 recipe.